\documentclass[a4paper,11pt]{article}
\pdfoutput=1 

\usepackage{jheppub} 

\usepackage[T1]{fontenc} 

\usepackage{graphics} 
  
\usepackage{xcolor}
\usepackage{epsfig} 
\input{epsf}

\usepackage{natbib}
\setcitestyle{square, numbers, comma, sort&compress}

\def\nn{\nonumber}
\def\be{\begin{equation}}
\def\ee{\end{equation}}
\def\bea{\begin{eqnarray}}
\def\eea{\end{eqnarray}}
\def\ms{\overline {\rm MS}}
\def\lm{L_m}
\def\leta{L_\eta}

\title{\boldmath Critical behavior of the 2d scalar theory: resumming the ${\rm N}^8{\rm LO}$ perturbative mass gap}

\author{Gustavo O. Heymans} 
\author{and Marcus Benghi Pinto}

\affiliation{Departamento de F\'{\i}sica, Universidade Federal de Santa
  Catarina, 88040-900 Florian\'{o}polis, SC, Brazil}

\emailAdd{gustavo.olegario@posgrad.ufsc.br}
\emailAdd{marcus.benghi@ufsc.br}

\abstract{We apply the optimized perturbation theory  (OPT) to resum the perturbative series describing the mass gap  of the bidimensional $\phi^4$  theory in the $\mathbb{Z}_2$  symmetric phase. Already at NLO  (one loop)   the method is capable of generating  a quite reasonable non-perturbative  result for the critical coupling.  At order-$g^7$ we obtain $g_c = 2.779(25)$ which compares very well with the state of the art ${\rm N}^8{\rm LO}$ result, $g_c = 2.807(34)$. As a novelty we    investigate  the supercritical region showing that it  contains some useful complimentary information  that can be used in  extrapolations to arbitrarily high orders.  } 
 
\begin{document} 
\maketitle
\flushbottom
\section{Introduction} 

The bidimensional scalar $\phi^4$ model describes a simple non integrable super-renormalizable theory which displays  rich phase transition patterns. When the original mass parameter, $m^2$, is positive the model has a mass gap and remains  invariant under the   $\mathbb{Z}_2$ transformation $\phi \to -\phi$  as far as one remains within the weak coupling regime. As the coupling ($g$) increases the mass gap decreases until the symmetry gets ultimately broken through a second order phase transition when a critical value, $g_c$, is attained \cite{chang,simon}. The case $m^2 <0$ displays an even richer phase transition structure in which the $\mathbb{Z}_2$ symmetry that is broken for $0 < g < {\tilde g}_c$ gets restored at $g= {\tilde g}_c$ through a second order transition \cite{serone2}. As the coupling further increases the model returns to the broken phase at $g =  {\tilde g}^\prime_c$. An interesting  duality between the   $\mathbb{Z}_2$ broken and unbroken phases, which allows to relate the three different critical couplings,  was discovered by Chang \cite{chang}.   Since its $\beta$ function vanishes at all orders in perturbation theory the model  represents a conformal field theory  at the critical coupling where  it becomes gapless. It then lies within the same universality class as the bidimensional Ising model.  These physically appealing characteristics  combined with its simplicity suggest that the strongly coupled bidimensional $ \phi^4$ model provides an excellent framework to test how accurately different non-perturbative techniques describe critical parameters associated with the phase transitions.  Indeed, a survey of the literature reveals that methods such as lattice simulations \cite{lattice1, lattice2, lattice3, lattice4},  Hamiltonian truncations (HT) \cite{hamilton1, hamilton2, hamilton3,hamilton4, hamilton5, hamilton6, hamilton7, hamilton8, hamilton9, hamilton10} as well as other resummation schemes \cite {paul} have  been recently used  to  determine the numerical value of $g_c$ for the symmetric ($m^2 > 0$) case. In some of the most recent investigations \cite {serone1,serone3} the self energy contributions to the physical mass have been perturbatively evaluated up to the ${\rm N}^8{\rm LO}$ before being Borel resummed to yield $g_c = 2.807(34)$. The availability of such a perturbative series provides us with an excellent opportunity to test  alternative resummation techniques as those prescribed by variational methods such as the optimized perturbation theory (OPT) \cite {opt_phi4, opt_qcd} to be considered here. Let us point out that similar 
approximations appear under  acronyms such as LDE (linear $\delta$ expansion \cite{lde}), VPT
(variational perturbation theory \cite {vpt}), and SPT (screened perturbation theory \cite{spt}).

The aim of most of these   variational approximations  is   to resum an originally perturbative series by combining the easiness of perturbative evaluations with some optimization criterion in order to produce non-perturbative results. In this way  the formal evaluations at each order generally involve only a handful of contributions which is certainly advantageous. A welcome feature is that the  renormalization program  can then be implemented by following the perturbative approach discussed in most standard textbooks {\it before} non-perturbative  results be produced through optimization. One way to implement this kind of approximation is to deform the original theory by a adding and subtracting a Gaussian term written terms of an arbitrary variational (mass) parameter, $\eta$, which represents a Lagrange multiplier. For example, in the case of the  $\phi^4$ scalar  theory to be considered here one deforms the original theory by shifting  the harmonic term $m^2 \phi^2 \to m^2\phi^2 + (1-\delta) \eta^2 \phi^2$  while multiplying the original couplings by a dummy   bookeeping parameter, $\delta$. Then, as one can easily check, the deformed Lagrangian density  interpolates between a free theory ($\delta =0$) and the original interacting case ($\delta =1$). Next, a physical quantity, $\Phi$, is evaluated {\it perturbatively } in powers of $\delta$ up to a given order $k$ producing $\Phi^{(k)}(\delta,\eta)$. One then sets $\delta=1$ (the original value) and fixes the optimum $\overline \eta$ by requiring that it satisfies the so called principle of minimal sensitivity (PMS), $(\partial \Phi)/(\partial \eta)|_{\overline \eta} =0$ \cite{pms1,pms2}.
In most situations this variational criterion produces non-perturbative results since the optimal $\overline \eta$ turns out to depend on the couplings in a non trivial way (often, via self consistent relations). The OPT method is known for exactly reproducing large-$N$ results already at the first non trivial order in many relevant situations \cite{npb}. Finite $N$ non-perturbative corrections are easily taken into account by considering few higher loop contributions. Another advantage concerns the case of an originally massless theories for in this case the variational mass also acts as an infra red regulator. This approximation has been successfully employed in many different physical situations involving symmetry breaking and phase transitions. In particular, applications related to condensed matter physics have shown that by including finite $N$ corrections in a  non-perturbative fashion the method was able to produce very accurate results regarding the critical dopant concentration in polyacetylene \cite {poly},  the critical temperature for homogeneous Bose gases \cite {bec}, and the phase diagram of magnetized planar fermionic systems \cite {GNmag}. Applications to high energy physics include the evaluation of quark susceptibilities \cite{tulio} and the phase diagram of effective QCD  models \cite {2CEP} among others.  Regarding gauge theories at finite temperatures and non vanishing baryonic densities a variant of the method, with renormalization group  properties, has been recently applied to determine the QCD equation of state for  dense hadronic matter at $T=0$ \cite {coldQCD} and $T\ne 0$ \cite {letterHOT, longHOT}. This variant, which has been originally dubbed renormalization group optimized perturbation theory (RGOPT) \cite{JLGN,JLalphas}, has produced results which are in excellent agreement with the state of the art lattice QCD predictions \cite {letterHOT, longHOT}.

In the present work the OPT capabilities at strong couplings will be tested   primarily to evaluate $g_c$ for the case $m^2 >0$. With this aim we shall resum the ${\rm N^8 LO}$ mass gap perturbative series which became recently available  \cite{serone3}.  Our other  goal is to investigate if some extra useful information can be acquired by exploring the supercritical region. The work is organized as follows. In the next section we review the  ${\rm N}^8{\rm LO}$ mass gap perturbative series  presented in Ref. \cite{serone3}. In Sec. III the OPT method is illustrated with an application to the mass gap at the two first non trivial orders. Numerical results for $g_c$ at  ${\rm N}^8{\rm LO}$ are obtained in Sec. IV. Then, in Sec. V, we explore the region where $g>g_c$.  Finally, our conclusions  are presented in Sec. VI.

\section{Reviewing the   perturbative mass gap series}

The well known $\phi^4$ model is described by the  following $\mathbb{Z}_2$ invariant Lagrangian density 
\be\label{eq:Lagrangian}
\mathcal{L} = \frac{1}{2}\partial_{\mu}\phi\partial^{\mu}\phi + \frac{m^2}{2}\phi^2 + \lambda\phi^4 \,.
\ee
In 2d, where the theory is super-renormalizable, $\lambda$   has canonical dimensions $[\lambda] = 2$. In this particular case the coupling  is finite so that $\beta_\lambda =0$ at all perturbative orders. Regarding the two point function the only  primitive divergence stems from tadpole (direct) contributions which do not depend on the external momenta. Hence, no wave renormalization is needed. 

Let us start by reviewing the perturbative evaluation of the physical mass squared. At the lowest perturbative order, an explicit evaluation using dimensional regularization  in the $\ms$ scheme gives
\be
M_{PT}^2 = m^2 + \lambda \frac{3}{\pi} \left ( \frac  {1}{\epsilon} + \lm \right ) + m_{ct}^2\;,
\label{MassDIV}
\ee
where we have defined
\be\label{eq:Lm}
 L_m \equiv \ln{\frac{\mu^2}{m^2}} \,,
\ee
with $\mu$ representing an arbitrary energy scale. Within the $\ms$ renormalization scheme only the pole  is eliminated by the counterterm, $m^2_{ct}$, so that the renormalized physical mass squared reads
\be
M_{PT}^2 = m^2 + \lambda \frac{3}{\pi}  \lm \,.
\ee
Requiring this quantity to satisfy the Callan-Symanzik equation
\be
(\mu \partial_\mu + \beta_{m^2} \partial_{m^2}) M^2 = 0\,,
\ee
one can fix the mass anomalous dimension to \cite {serone2}
\be\label{betam}
\beta_{m^2} = -\lambda \frac{3}{\pi} \,.
\ee
Therefore, 
\be
m^2(\mu) = m^2(\mu_0) - \lambda \frac{3}{\pi}\ln{\frac{\mu^2}{\mu_0^2}} \,,
\ee
where $\mu_0$ represents a reference scale.
As already emphasized the only primitive divergence associated with the two point function is the one which appears in Eq. (\ref {MassDIV}). This means that by consistently considering the counterterm $m_{ct}^2 = -  \lambda   3 / (\pi \epsilon)$ the physical mass squared will remain finite at any perturbative order. Note  also that $\beta_{m^2}$, as given by Eq. (\ref{betam}), remains valid as higher orders are considered.  In the case where tadpoles are present the finite  perturbative physical  mass squared at ${\cal O}(\lambda^8)$ reads \cite{serone3}
\bea\label{eq:MassO(8)}
M^2_{PT} &=& m^2 + \frac{3\lambda}{\pi}\lm - \frac{9\lambda^2}{\pi^2m^2}\lm - \frac{3\lambda^2}{2m^2}\nn\\
& +& \frac{\lambda^3}{(m^2)^2}\Bigg\{ \frac{9}{\pi} + \frac{63}{2\pi^3}\zeta(3) + \frac{27}{\pi^3}\lm  + \frac{9}{2\pi}\lm + \frac{27}{2\pi^3}\lm^2 \Bigg\} \nn \\
&-& \frac{\lambda^4}{(m^2)^3}\Bigg\{14.655869(22) + \left(6 + 5\pi^2 + 14\zeta(3)\right)\lm +  \frac{27}{2\pi^4}\left(9 + \pi^2\right)\lm^2 + \frac{27}{\pi^4}\lm^3 \Bigg\} \nn \\
&+& \frac{\lambda^5}{(m^2)^4} \Bigg\{65.97308(43) + 51.538171(63)\lm + \frac{81}{4\pi^5}\left(36 + 17\pi^2 + 42\zeta(3)\right)\lm^2  \nn \\ 
&+& \frac{81}{2\pi^5}\left(11 + \pi^2 \right)\lm^3 + \frac{243}{4\pi^5}\lm^4 \Bigg\} \nn \\
 &-& \frac{\lambda^6}{(m^2)^5} \Bigg\{347.8881(28) + 301.2139(16)\lm + 114.49791(12)\lm^2 \nn \\ 
 &+& \frac{81}{2\pi^6}\left(105 + 37\pi^2 + 84\zeta(3) \right)\lm^3 +  \frac{243}{4\pi^6}\left(25 + 2\pi^2 \right)\lm^4 + \frac{729}{5\pi^6}\lm^5 \Bigg\} \nn \\
 &+& \frac{\lambda^7}{(m^2)^6}\Bigg\{2077.703(36) + 1948.682(14)\lm +  828.4327(39)\lm^2 +  205.20516(19)\lm^3  \nn \\
 &+& \frac{243}{8\pi^7}\left(675 +197\pi^2 + 420\zeta(3)\right)\lm^4 + \frac{729}{5\pi^5}\left(137 + 10\pi^2\right)\lm^5 + \frac{729}{5\pi^7}\lm^6 \Bigg\} \nn \\
 &-& \frac{\lambda^8}{(m^2)^7}\Bigg\{13771.04(54) + 13765.22(21)\lm + 6373.657(40) \lm^2 + 1778.1465(75)\lm^3 \nn \\
 &+& 323.93839(27)\lm^4 +  \frac{2187}{20\pi^8}\left(812 + 207\pi^2 + 420\zeta(3) \right)\lm^5 \nn \\ &+&\frac{2187}{20\pi^8}\left(147 + 10\pi^2\right)\lm^6 + \frac{6561}{7\pi^8}\lm^7 \Bigg\}\,. 
 \label{Mpt}
\eea
Note that by setting $\mu=m$ all $\lm$ dependent (tadpole) terms appearing in Eq. (\ref{Mpt}) vanish and one retrieves the series considered in Ref. \cite {serone1}. However, as we shall explicitly see,  these terms should not be discarded prior to implementing the OPT mass shift.  The reason is that after expanding in powers of $\delta$ the $\lm$ terms become $\ln [\mu^2/(m^2+\eta^2)]$   eventually giving non vanishing contributions even if one later chooses $\mu=m$. 

\section {OPT resummation }

\begin{figure}[ht!]

    \centering
    \includegraphics[scale=.20]{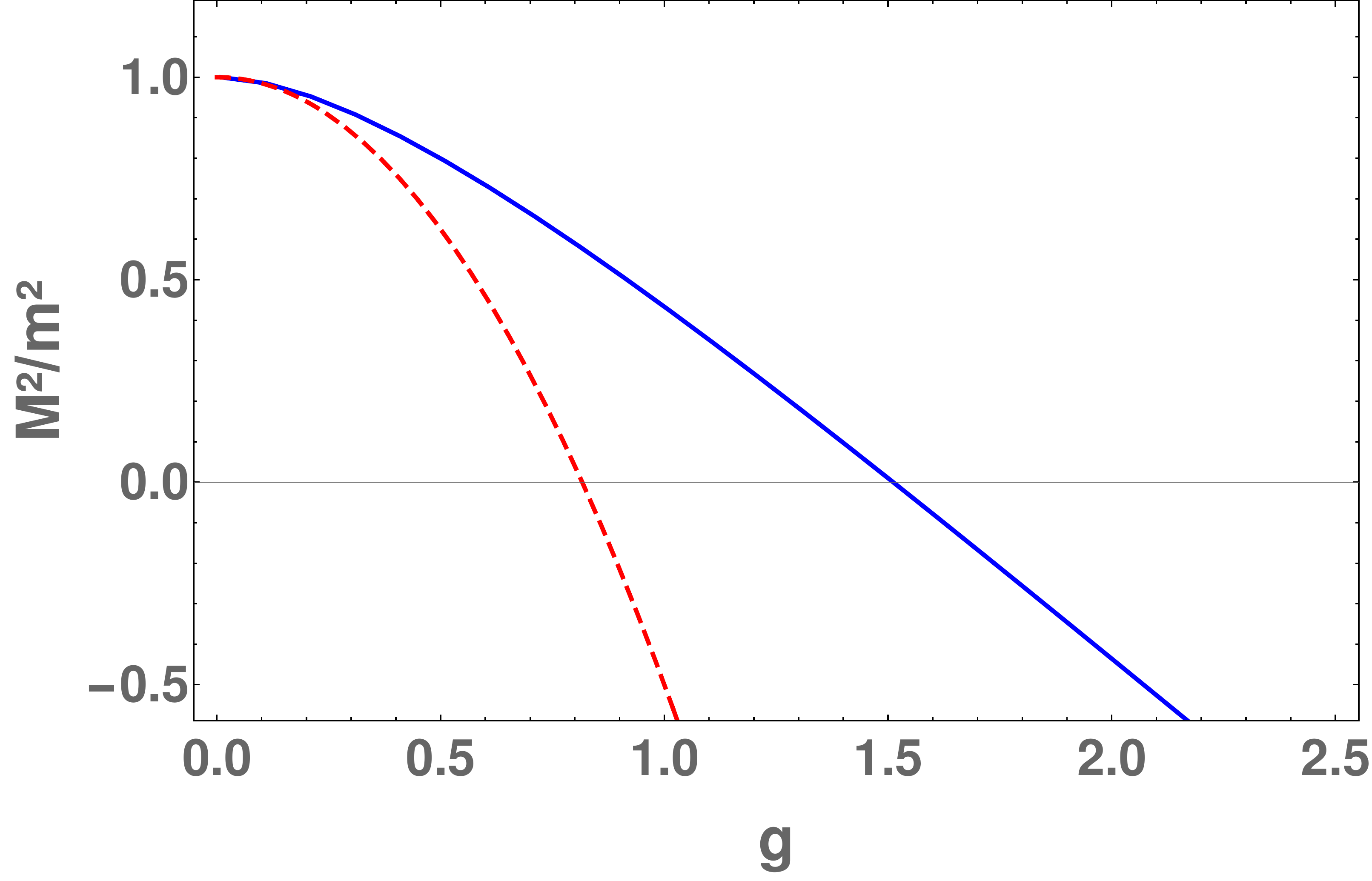}
        \caption{The physical mass squared, in units of $m^2$,  as a function of $g$ obtained with standard PT (dashed line) and OPT (continuous line) at the two loop level. The critical couplings occur at $g_c = \sqrt{2/3} = 0.82$ (PT) and $g_c = 1.511$ (OPT). }
    \label{Fig1}
\end{figure}

Now, to implement the OPT procedure one considers Eq. (\ref {eq:MassO(8)}) with the following  replacements \cite{opt_phi4}
\bea\label{eq:OPT}
m^2(\mu)= m^2(\mu_0) -   \frac{3\lambda}{\pi}\ln{\frac{\mu^2}{\mu_0^2}} &\longrightarrow& m^2(\mu) + \eta^2(1-\delta) = m^2(\mu_0) + \eta^2(1-\delta)- \delta  \frac{3\lambda}{\pi}\ln{\frac{\mu^2}{\mu_0^2}} \nn \\ \nn
\lambda &\longrightarrow& \delta\lambda \,,\\
\label{replaceOPT}
\eea
where $\mu_0$ is a reference scale. Next,  by reexpanding to a given order $\delta^{(k)}$ one obtains the OPT physical mass squared, $M^2$. To understand how the method works it is convenient to extract the maximum of information in an analytical fashion. This can be achieved by considering the first two non trivial lowest order  contributions given by
\bea\label{eq:MassO(2)}
M^2(\mu)& = & m^2(\mu_0)  - \delta  \frac{3\lambda}{\pi}\ln{\frac{\mu^2}{\mu_0^2}} + \eta^2 (1- \delta) + \delta \frac{3\lambda}{\pi} \ln{\frac{\mu^2}{[m(\mu_0)^2+\eta^2]}}+ \delta^2  \frac{3\lambda \eta^2}{\pi[m(\mu_0)^2+\eta^2]}  \nn \\
&-& \delta^2 \frac{9\lambda^2}{\pi^2[m(\mu_0)^2+\eta^2]}\ln{\frac{\mu^2}{[m(\mu_0)^2+\eta^2]}}+\delta^2 \frac{9\lambda^2}{\pi^2[m(\mu_0)^2+\eta^2]}\ln \frac{\mu^2}{\mu_0^2}
\nn \\
&-& \delta^2 \frac{3\lambda^2}{2[m(\mu_0)^2+\eta^2]} + {\cal O}(\delta^3) \;. 
\label{M2mus}
\eea
 It is now a trivial matter to rearrange the logarithms to see that at any arbitrary scale, $\mu^\prime$, the  OPT mass can be written as\footnote {As in the purely perturbative case this feature remains valid as higher order contributions are considered.} $M^2(\mu^\prime) = M^2(\mu_0)$ where
\bea\label{eq:MassO(2)RG}
M^2(\mu_0) &=& m^2(\mu_0)   + \eta^2 (1- \delta) + \delta \frac{3\lambda}{\pi}\leta + \delta^2  \frac{3\lambda \eta^2}{\pi[m(\mu_0)^2+\eta^2]} - \delta^2 \frac{9\lambda^2}{\pi^2[m(\mu_0)^2+\eta^2]}\leta \nn \\
&-&  \delta^2 \frac{3\lambda^2}{2[m(\mu_0)^2+\eta^2]} + {\cal O}(\delta^3) \,.
\label{m2mu0}
\eea
The following definition has been used in the previous relation 
\be\label{eq:Leta}
 L_\eta \equiv \ln{\frac{\mu_0^2}{[m(\mu_0)^2+\eta^2]}} \,.
\ee
Having understood that the complete standard perturbative renormalization procedure is not spoiled by the OPT simple replacements we can turn to the optimization procedure. With this aim let us set $ \mu_0 = m(\mu_0)$ in Eq. (\ref {m2mu0}) so that our results can be directly compared to those of Ref. \cite {serone1} (or, equivalently, to those presented in Ref. \cite {serone3}  for the particular  case $\kappa = \lm \equiv 0$). Next, let us define the dimensionless coupling
\be
g\equiv \frac{\lambda}{m^2} \;.
\ee
Then, in units of $m^2$ the physical mass squared becomes
\be\label{eq:MassO(2)dimless}
\frac{M^2}{m^2}= 1   + \frac{\eta^2}{m^2} (1- \delta) + \delta \frac{3g}{\pi}\leta +  \delta^2 \frac{1}{(1+\eta^2/m^2)} \left [g\frac{3 \eta^2}{\pi m^2} - g^2 \frac{9}{\pi^2} \leta - g^2 \frac{3}{2}   \right ]  + {\cal O}(\delta^3)\;.
\ee
At order-$\delta$  the optimal solution is just ${\overline \eta}=0$ implying that the trivial perturbative solution $M^2/m^2 =1$ is recovered. A non trivial result is obtained at order-$\delta^2$ where, after setting $\delta=1$ and applying the variational criterion \cite{pms1,pms2} 
\begin{equation}
\frac{\partial M^2}{\partial \eta} \Bigr |_{\overline \eta} = 0 \;,
\label{pms}
\end{equation}
to Eq. (\ref {eq:MassO(2)dimless}), one obtains as  solutions the trivial $\overline \eta =0$ as well as  the highly non-perturbative relation
\begin{equation}
{\overline \eta}^2 = m^2 \left [ \frac{3 g}{\pi} W\left ( \frac{\pi \exp[1 + \pi/{(3 g)} + \pi^2/6]} {3g }  \right ) - 1 \right ]\,,
\label {Nontrivial}
\end{equation}
where $W$ represents the Lambert-$W$ function.  Remark that exactly at $g=0$ the non trivial solution, Eq. (\ref {Nontrivial}), would  give ${\overline \eta}^2 = - m^2$ leading to divergences but obviously in this case one has a free theory and the optimal mass is just ${\overline \eta} =0$. Fig. \ref {Fig1} compares the OPT result for  $M^2(g)$ with the standard PT prediction at order-$g^2$ (two loop level) showing that a second order quantum transition takes place when $M^2(g_c)=0$. 

Before pushing the evaluation of $M^2$ to higher orders it is important to remark that the authors of Ref. \cite {serone3} noticed that by resumming the series for $M$, instead of the one for $M^2$, much better results could be obtained due to the fact that $M$ approaches the critical point smoothly. Therefore, for comparison purposes, we shall also resum this quantity to determine $g_c$ from the condition $M(g_c)=0$. After taking the square root of Eq. (\ref{Mpt}), reexpanding to order $\lambda^8$, and carrying out the OPT replacements represented by Eqs. (\ref {replaceOPT}) one easily obtains  the OPT series up to order-$\delta^8$. As we did in the $M^2$ case let us start by considering $M$ expanded to order-$\delta^2$. The explicit result reads
\begin{eqnarray}
\frac{M}{m} &= &\sqrt{1 + \eta^2/m^2} - \delta\frac{1}{2\sqrt{1 + \eta^2/m^2}}\left ( \frac {\eta^2}{m^2} - g \frac{3}{\pi} L_\eta \right )\nn \\
& -&  \delta^2 \frac{\eta^2}{m^2(1 + \eta^2/m^2)^{3/2}}\left [ \frac{\eta^2}{8m^2} - g \frac{1 }{2\pi }\left ( 3 + \frac{1}{2} L_\eta\right )\right ]\nn \\
&-& \delta^2 g^2 \frac{3 }{(1 + \eta^2/m^2)^{3/2}} \left [   \frac{1}{4} +  \frac{3}{2\pi^2}L_\eta \left (1  +  \frac{1}{4} L_\eta \right ) \right ]
+ {\cal O}(\delta^3) \;.
\label{M1}
\end{eqnarray}
After setting $\delta=1$ and applying the PMS condition one immediately notices that, contrary to what happens when resumming $M^2$, optimizing $M$  at order-$\delta$ yields ${\overline \eta} =0$ together with the non trivial result
\begin{equation}
{\overline \eta}^2 = m^2 \left [\frac{3g}{\pi}  W\left ( \frac{\pi \exp[2 + \pi/(3g)] }{3g} \right ) - 1 \right ] \;,
\label{etabar1}
\end{equation}
which is rather similar to the optimal mass obtained by resumming the squared mass at order-$\delta^2$. However, in opposition to what happens in the $M^2$ case, the optimization of $M$ does not furnish a non trivial real result at order-$\delta^2$. Then,  injecting   $\overline \eta$  as given by Eq. (\ref {etabar1}) back into Eq. (\ref {M1})  at ${\cal O}(\delta)$ yields $g_c =3.760$. Although numerically not so accurate (e.g., when compared to $g_c = 2.807(34)$ of Ref. \cite{serone1}), this is still a quite remarkable result given that it has been generated by resumming only a simple one loop contribution. This result  also emphasizes the crucial role  played by the OPT tadpole terms $\leta$ which,  in contrast to purely perturbative $\lm$,  survive the scale choice $\mu=m$.

\section{Numerical results at ${\rm N}^8{\rm LO}$}

\begin{figure}[ht!]
    \centering
    \includegraphics[scale=0.2]{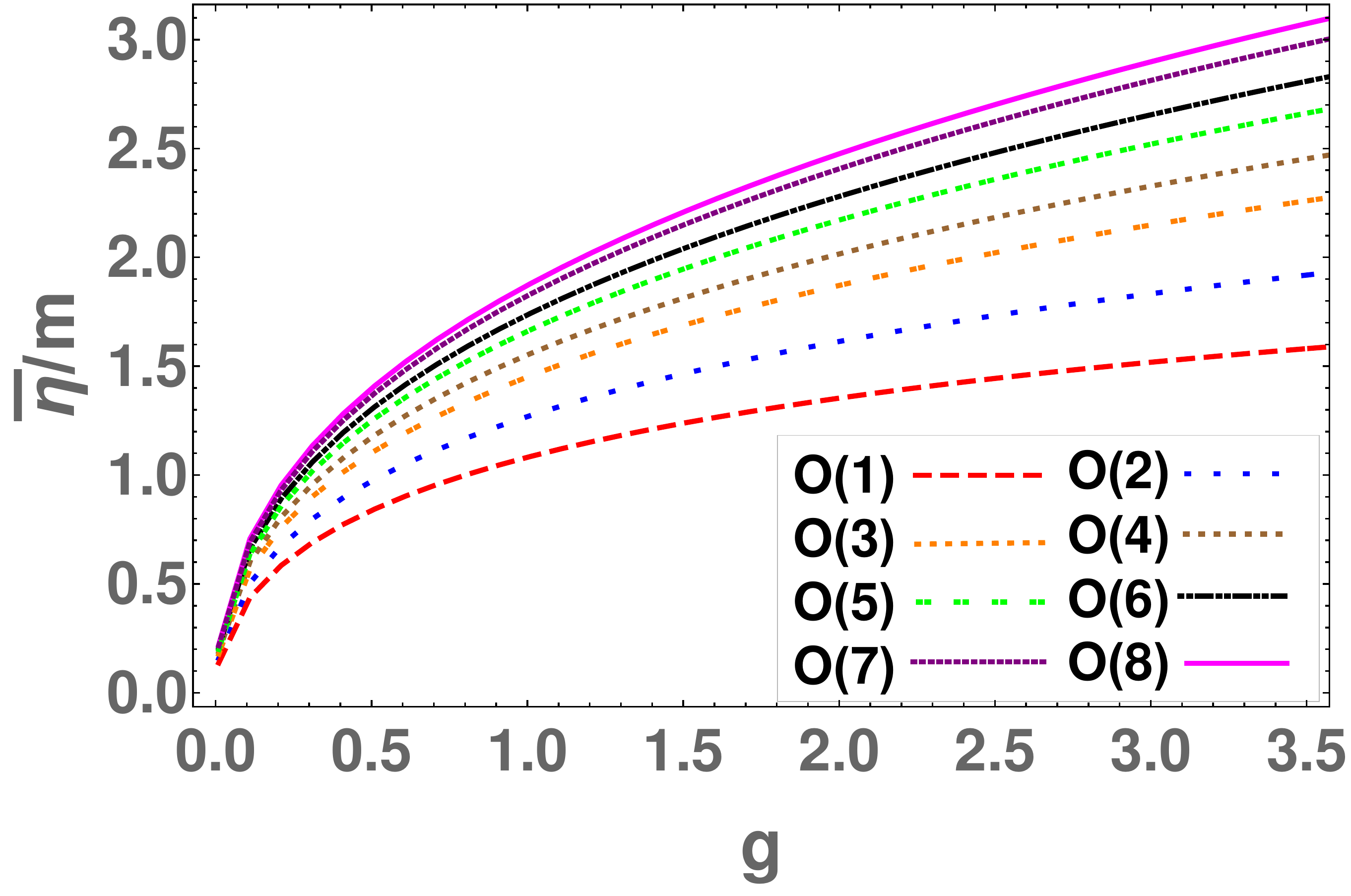}
    \caption{The optimized variational mass, in units of $m$, as a function of $g$ obtained by resumming $M$ (odd orders) and $M^2$ (even orders).}
    \label{Fig2}
\end{figure}

\begin{figure}[ht!]

    \centering
    \includegraphics[scale=0.2]{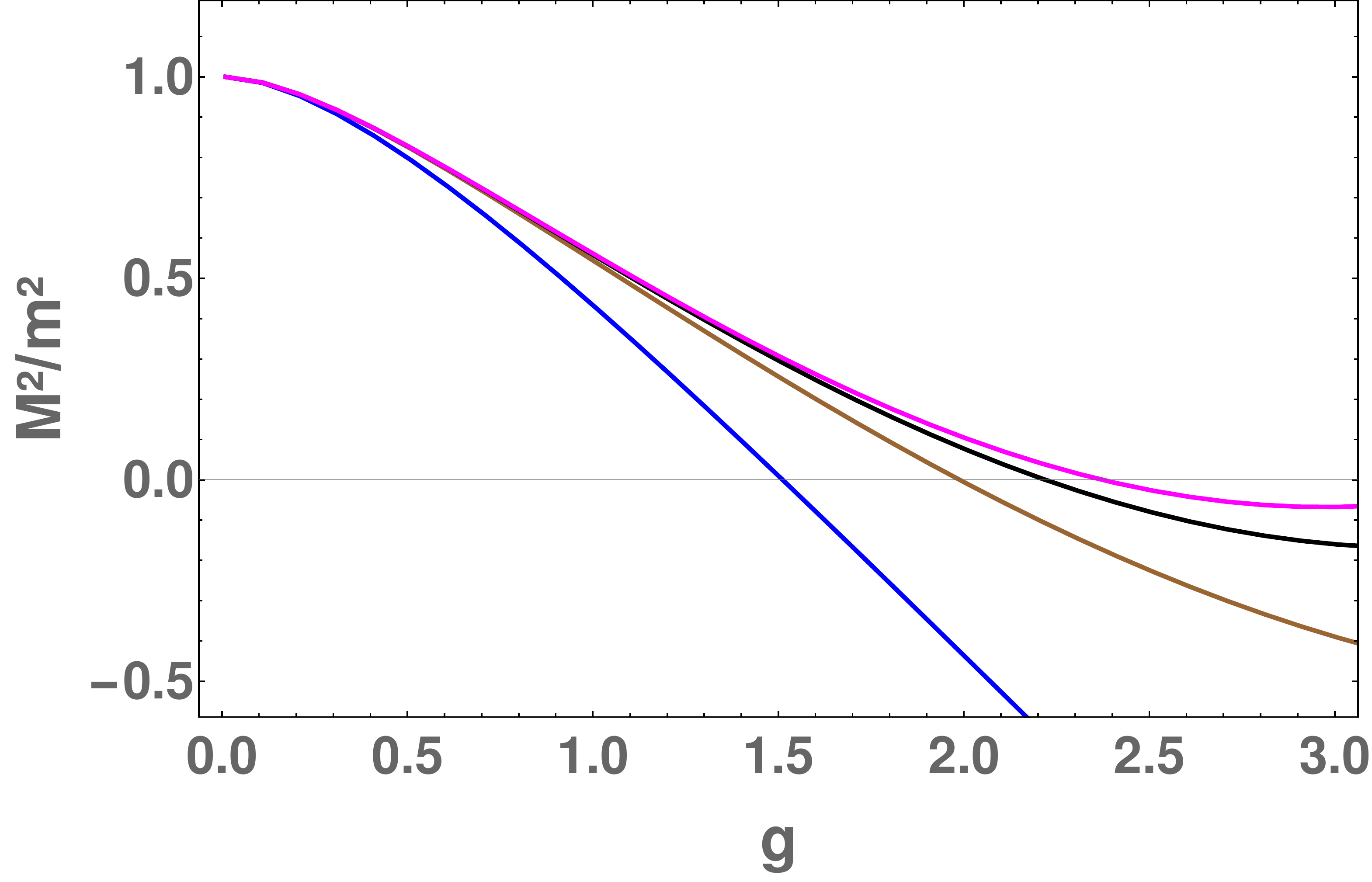}
    \qquad
     \includegraphics[scale=0.2]{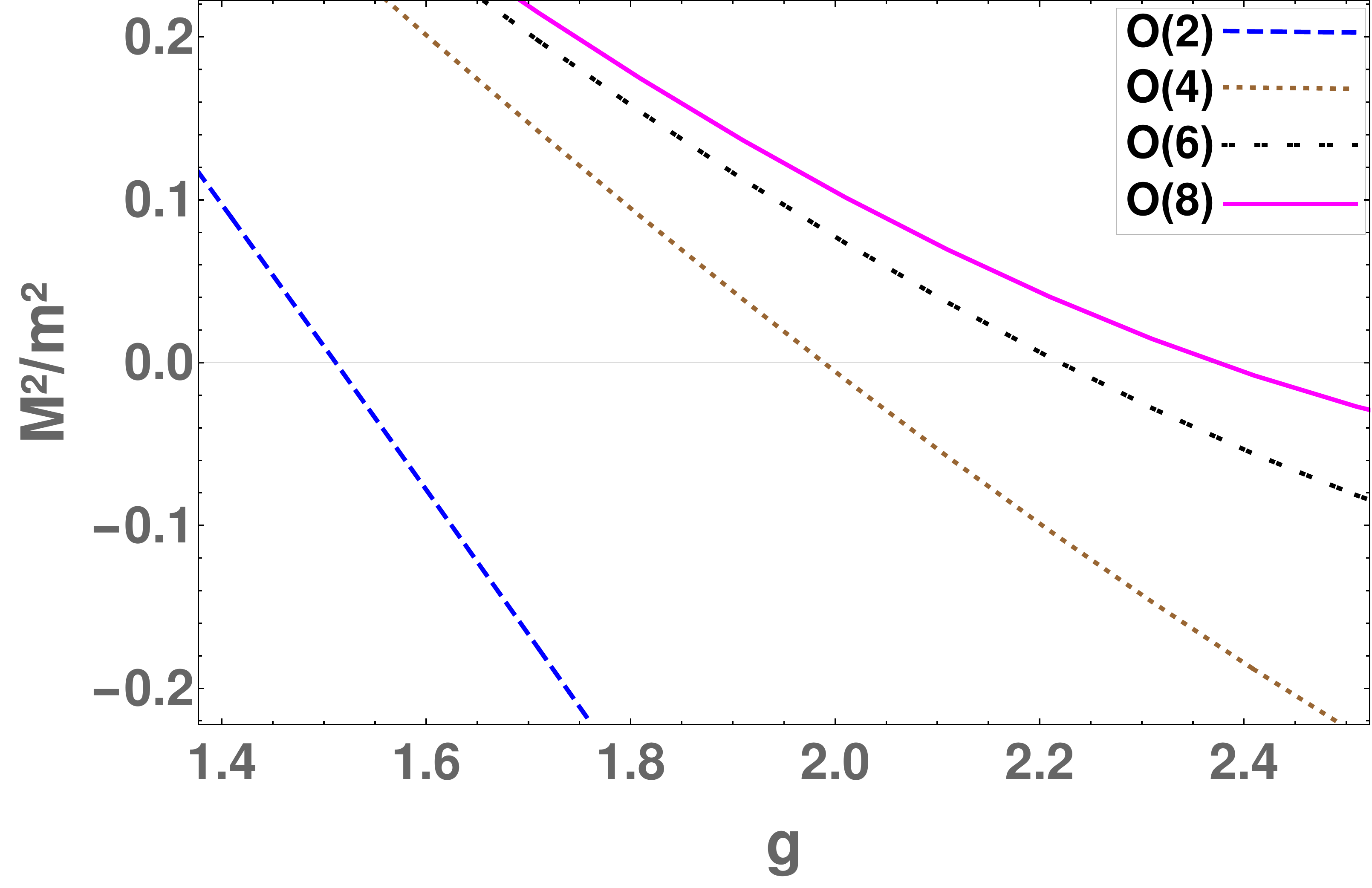}
    \caption{Left panel: The physical mass squared, in units of $m^2$,  as a function of $g$ at even orders. Right panel: Same as the left panel but zoomed. }
    \label{Fig3}
\end{figure}

Now, by following the steps which led to the order-$\delta^2$ mass gap result it is a simple matter to consider higher order corrections up to ${\cal O}(\delta^8)$ starting from  the perturbative result, Eq. (\ref {Mpt}). The whole procedure can be easily carried out in a numerical fashion. Exactly as it happened at NLO it turns out that real solutions  for $\overline \eta$ can only be obtained at even(odd) orders when $M^2$($M$) is optimized. The complete set of results for the optimal $\overline \eta$ is displayed in Fig. \ref {Fig2}. Using the results for  $\overline \eta$ at even orders one can then investigate the physical mass squared to determine the critical coupling from $M^2(g_c)=0$ as Fig. \ref{Fig3} illustrates.

\noindent

\begin{figure}[ht!]
    \centering
    \includegraphics[scale=0.18]{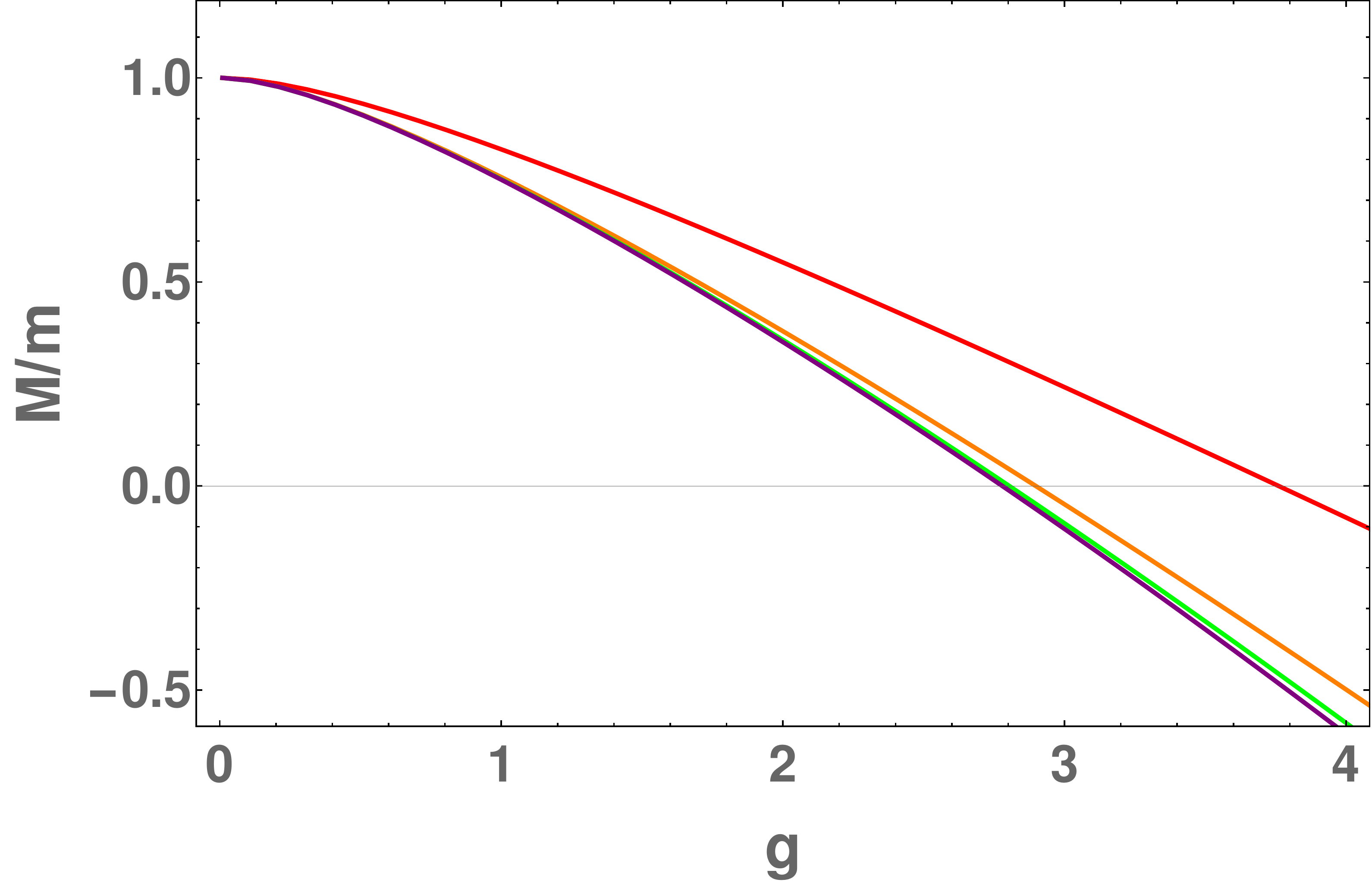}
    \qquad
     \includegraphics[scale=0.18]{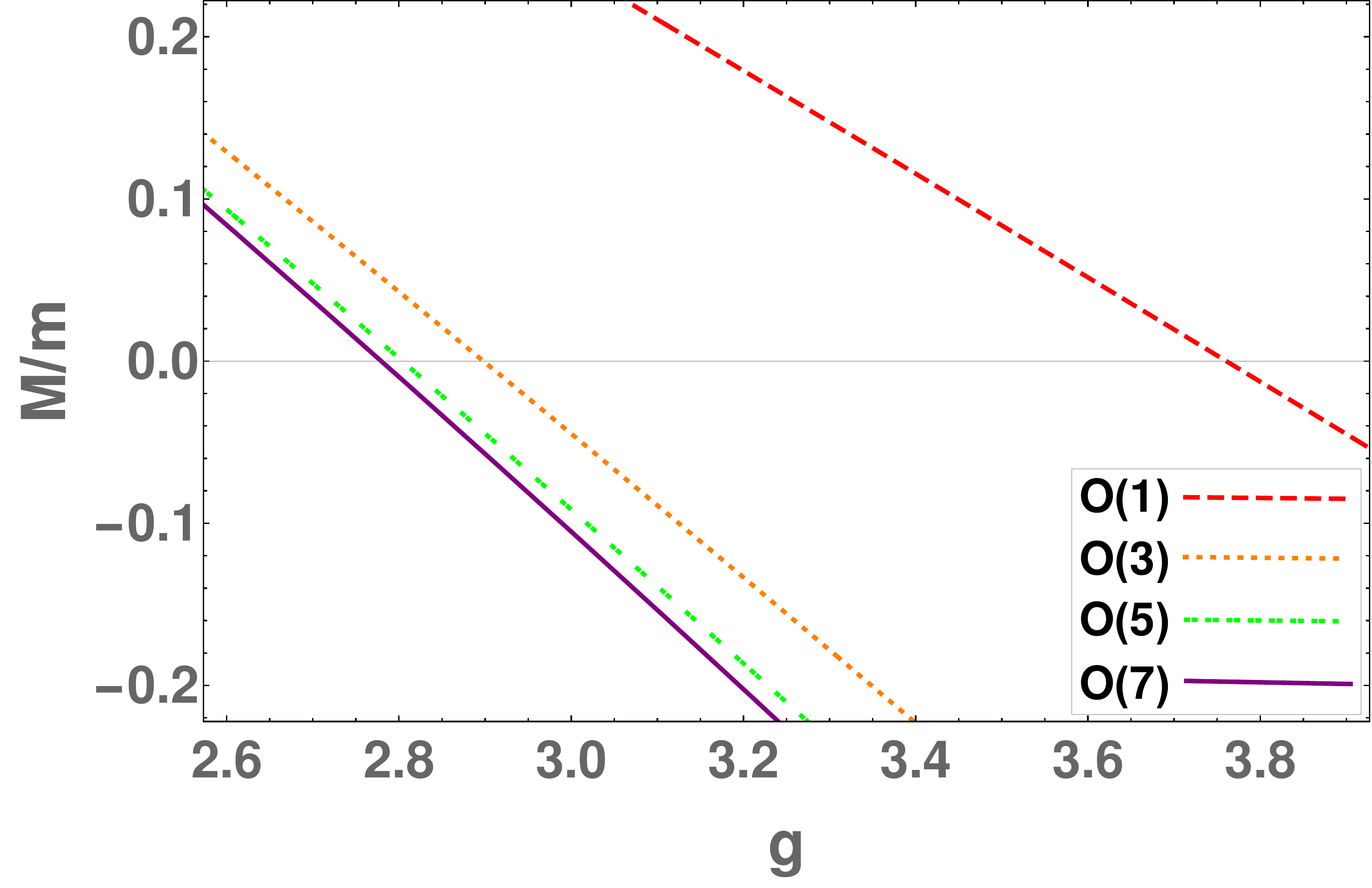}
    \caption{Left panel: The  physical mass, in units of $m$,  as a function of $g$ at odd orders. Right panel: Same as the left panel but zoomed.}
    \label{Fig4}
\end{figure}
At the same time, at odd orders the critical coupling value should be determined from the condition $M(g_c)=0$.  Fig. \ref{Fig4} shows $M(g)$  indicating that the value of $g_c$ decreases as more orders are considered in  opposition to what happens when $M^2$ is resummed. 
\begin{table}[ht!]
    \centering
    \begin{tabular}{ |c||c||c| }
        \hline
        $g$ & $M(g)$ &  Ref. \\
        \hline
        \hline
        {} &0.979733(5) & \cite{hamilton6,hamilton7}  \\ 
        0.02 &0.9797313(4) & \cite{serone1} \\
        {} &0.9797315(1) &  this work \\
        \hline
         {} &0.7494(2)& \cite{hamilton6,hamilton7}  \\ 
        1 &0.7507(5) & \cite{serone1} \\
        {} &0.750520(2) &  this work \\
        \hline
        {} &0.345(2) & \cite{hamilton6,hamilton7}  \\ 
        2 &0.357(5) & \cite{serone1} \\
        {} &0.352838(14) & this work \\
        \hline
    \end{tabular}
    \caption{Comparing some OPT values for $M(g)$, in units of $m$, with those predicted by Hamiltonian truncations \cite {hamilton6,hamilton7} and Borel resummation \cite{serone1}.  }
\label{tabela1}    
\end{table}

\begin{figure}[ht!]
    \centering
    \includegraphics[scale=0.22]{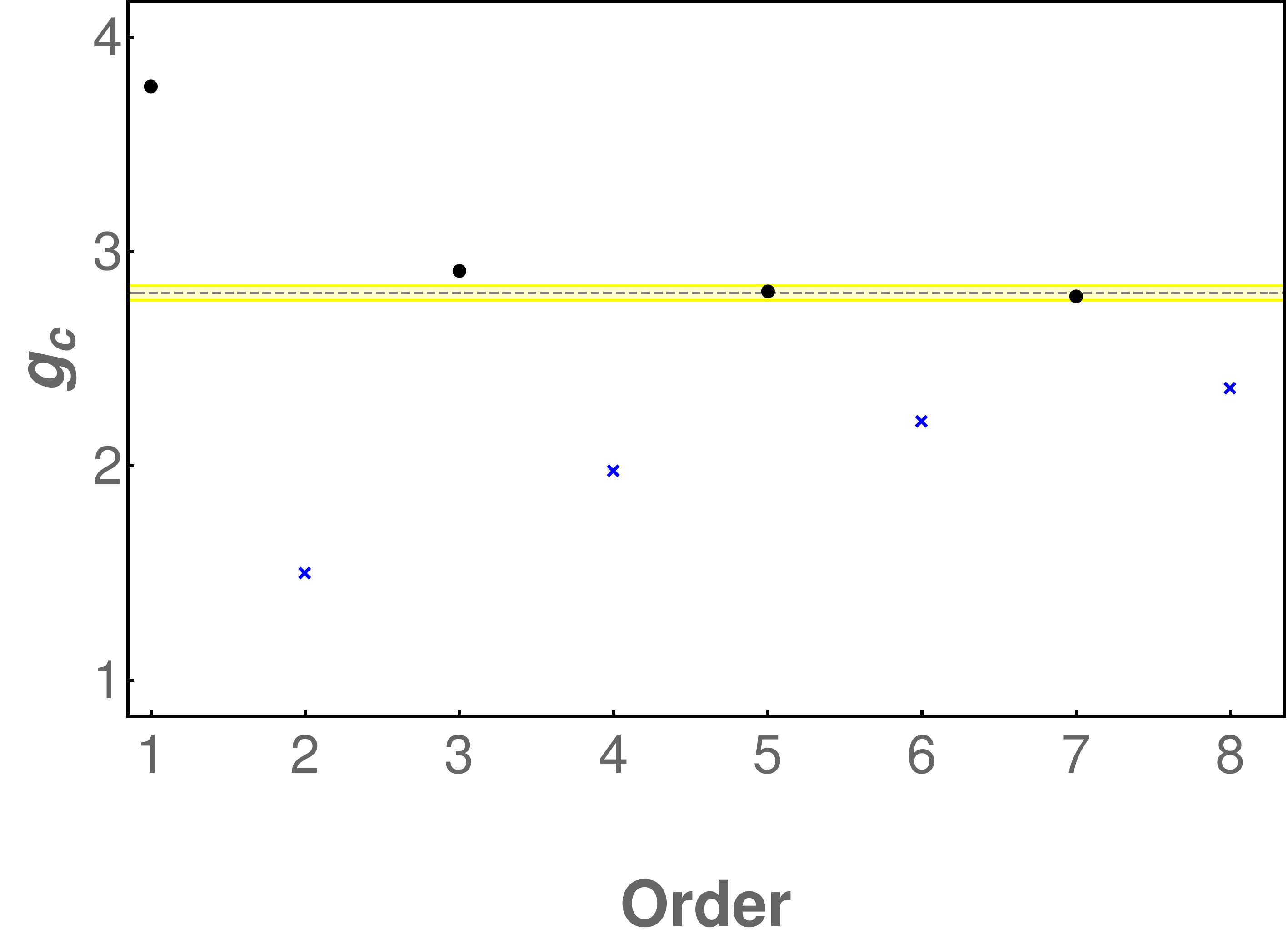}
    \caption{Order by order predictions for the critical coupling. The dots correspond to values obtained by resumming $M$ at odd orders while the crosses represent the values obtained by resumming $M^2$ at even  orders. The straight line represents the prediction  $g_c = 2.807(34)$ obtained in Ref. \cite {serone1} by resumming $M$ to ${\rm N}^8{\rm LO}$. }
    \label{Fig5}
 \end{figure}

\begin{table}[ht!]
    \centering
    \begin{tabular}{ |c|c||c|c| }
        \hline
        \multicolumn{2}{|c||}{Resummed $M$}&
        \multicolumn{2}{|c|}{Resummed $M^2$} \\
        \hline
        \hline
        order & $g_c$ & order & $g_c$\\
        \hline
        1 & 3.76015 & 2 & 1.51147  \\ 
        3 & 2.89809 & 4 & 1.98859  \\
        5 & 2.80400 & 6 & 2.21970  \\
        7 & 2.77947 & 8 & 2.37301 \\
        \hline
    \end{tabular}
    \caption{Critical couplings obtained from the OPT resummation of $M$($M^2$) at odd(even) orders.}
\label{tabela2}    
\end{table}
 
Table \ref{tabela1} compares the OPT predicted values for $M(g)$, at some representative coupling values,  with those furnished by Hamiltonian truncations   \cite {hamilton6,hamilton7} and Borel resummation  \cite{serone1}. The comparison indicates a  particularly good agreement between our results and those from Ref. \cite{serone1}.
As far as the critical coupling value is concerned  Fig. \ref {Fig5} compares the OPT predictions, at different $\delta$ orders, with the state of the art result obtained in Ref. \cite {serone1}. The figure clearly indicates that also within the OPT  resumming $M$ produces a much faster  convergence than  resumming $M^2$. The actual numerical values are presented in Table \ref{tabela2}. Not surprizingly, in comparison with $g_c = 2.807(34)$,  our most accurate  prediction happens at ${\rm N}^7{\rm LO}$ which represents the highest available odd order for which  $M$ can be optimized in the present work.

\section{Exploring the supercritical  region}

Having established the reliability of the OPT as an efficient resummation tool let us  now investigate what happens at coupling values greater than $g_c\equiv g_c^{(w)}$ which represents a (supercritical) region that has been seldom explored. Our motivation stems from a  very recent investigation \cite{serone4} in the three dimensional case where the authors observed that a second transition, leading back to the symmetric phase, could in principle occur at some coupling value, $g_c^{(s)}$, greater than $g_c^{(w)}$. This intriguing possibility (which in that work occurs within a particular renormalization scheme) has not been fully explored in Ref. \cite {serone4} because $g_c^{(s)}$ is reached after the theory has passed a phase transition. In this case, resummations of perturbative series  are not guaranteed to work since $g_c^{(w)}$ represents a non analytical point. Despite of that we shall see that by exploring this region  one gets  extra information that can be used to further improve convergence properties. To do that let us  consider once again the physical mass squared, $M^2$, since this quantity is often taken to represent  the order parameter in studies related to phase transitions in scalar models \cite {OPTphi4} while  $M$ does not have a clear physical interpretation  beyond $g_c$. Our NNLO result suggests that  after being broken at  $g_c^{(w)}$   the  $\mathbb{Z}_2$ symmetry could get restored at a higher value, $g_c^{(s)}$. Since at this first non trivial  order $g_c^{(s)} \simeq 14 g_c^{(w)}$ one could  dismiss the second transition on the grounds that it takes place at far too high couplings where the resummation is possibly  less effective as already emphasized. Within this point of view it would be reasonable to  expect that if indeed the  transition associated with $g_c^{(s)}$  emerges as an artifact of our approximation  the difference between $g_c^{(s)}$ and $g_c^{(w)}$ should further increase until $g_c^{(s)}$ eventually disappears, as higher orders are considered. 

\begin{table}[ht!]
    \centering
        \begin{tabular}{ |c|c|c|  }
            \hline
            Order  & $g_c^{(w)}$ & $g_c^{(s)}$ \\
            \hline
            2 & 1.51 & 21.4 \\ 
            4 & 1.98 & 5.89 \\
            6 & 2.21 & 4.16 \\
            8 & 2.37 & 3.53 \\ \hline
        \end{tabular}
    \caption{The first and second critical points, $g_c^{(w)}$ and $g_c^{(s)}$, from resumming $M^2$. }
 \label{tabela3}   
\end{table}
\begin{figure}[ht!]
    \centering
    \includegraphics[scale=0.18]{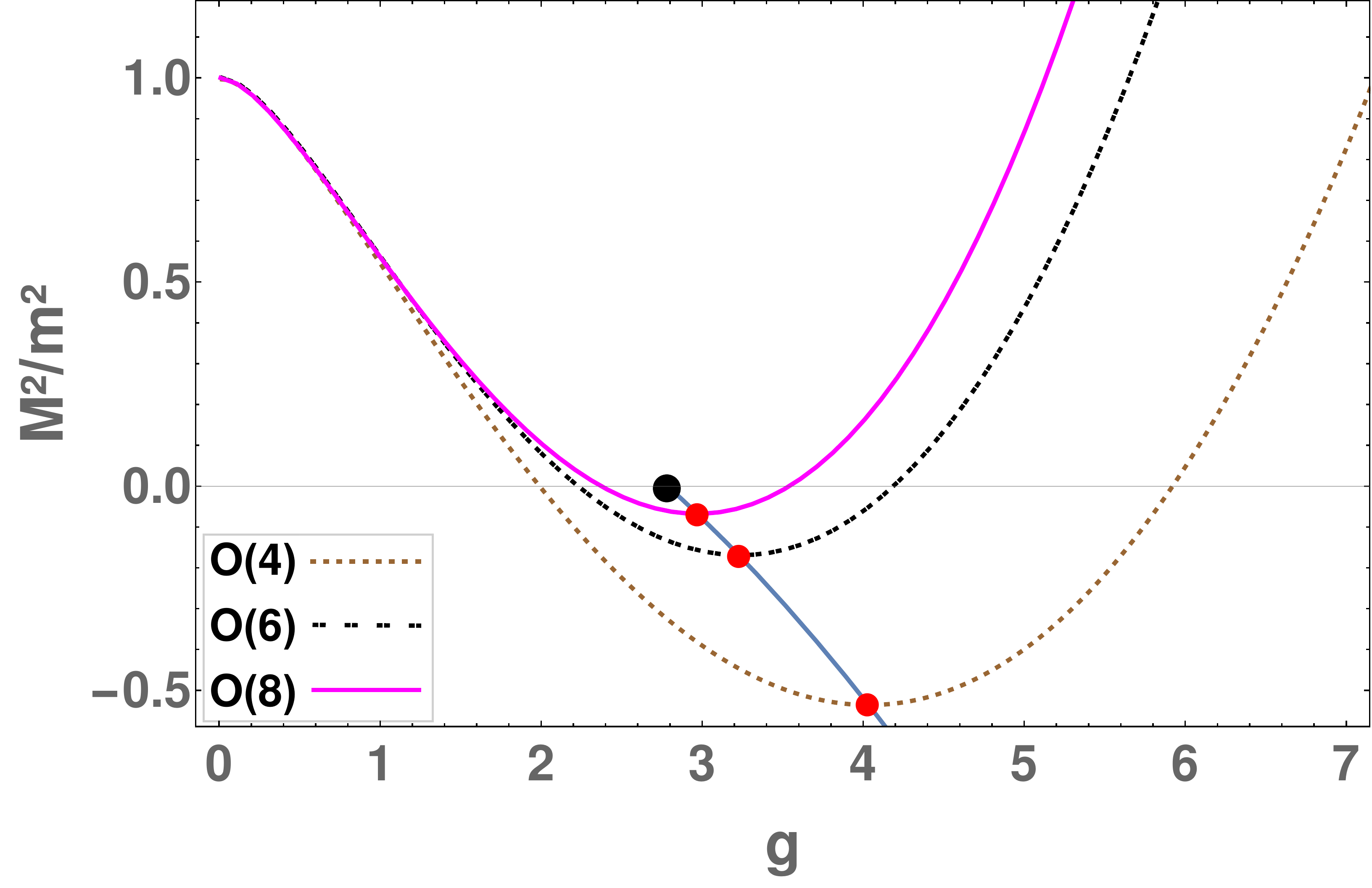}
    \qquad
     \includegraphics[scale=0.18]{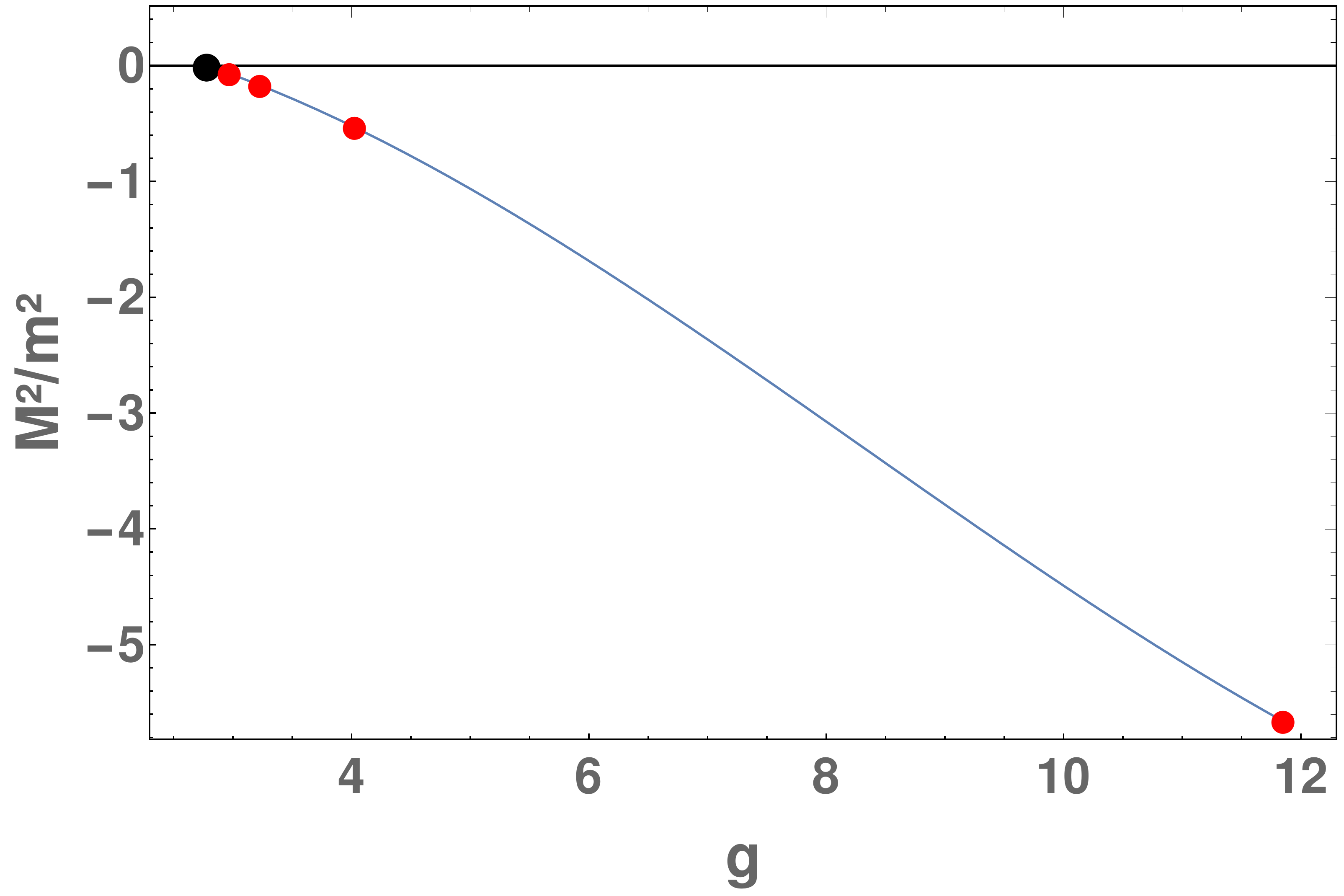}
    \caption{Left panel: The  physical mass squared, in units of $m^2$,  as a function of $g$ at even orders. For clarity the order-$g^2$ result is not shown in this panel. Right panel: Location of the minima from ${\cal O}(2)$ to ${\cal O}(8)$ (light dots).  The dark dot locates the extrapolated minimum corresponding to an arbitrarily high order.}
    \label{Fig6}
\end{figure}
However, table \ref{tabela3} reveals that at increasing orders the difference between the two critical coupling values decreases reaching  $g_c^{(s)} \simeq 1.5 g_c^{(w)}$ at ${\rm N}^8{\rm LO}$.  As a digression we note that, physically, the supercritical scenario observed in Fig. \ref{Fig6} reminds us of a reentrant phase which is a characteristic of many condensed matter systems.  One of the best known examples is the symmetry pattern observed in potassium sodium tartrate tetrahydrate most commonly known as the Rochelle salt which, as the temperature increases, goes   from a more symmetric orthorhombic crystalline structure to a less symmetric monoclinic structure at $T\simeq 255\, K$ \cite{ISB}. It then returns to be orthorhombic phase at $T\simeq 297 \,K$, till it melts at $T\simeq 348\,K$   therefore exhibiting  intermediary inverse symmetry breaking through a reentrant phase (see  Ref. \cite{condmatt} for a detailed list of different materials exhibiting this phenomenon).
Naively, Fig. \ref{Fig6} could then lead us to think that  a similar situation  takes place in the present bidimensional case where, instead of the temperature, the coupling drives the (quantum) transition.   However, although physically interesting, this exotic phase transition pattern is unlikely to emerge within the bidimensional $\phi^4$ model since our present investigation of $M$, at odd orders, indicates only one value at which $M(g_c)=0$. Moreover, the existence of a unique critical coupling  is also supported by other robust investigations \cite {hamilton6, hamilton7, serone1, serone3}.  Then, in the light of these results, we shall  not assume that a  reentrant transition  really takes place within the simple theory being studied here (in opposition to what happens in some $O(N)\times O(N)$ scalar models \cite{julia}). 
Nevertheless, even when this pragmatic view is adopted the supercritical region may still offer  interesting possibilities regarding high order extrapolations.  To see that let us first recall that, as seen in the previous section, resumming $M$ seems to be more effective than resumming $M^2$.  Next, let us refer again to Fig. \ref {Fig6} to point out that each curve presents a minimum lying within the $g_c^{(w)}-g_c^{(s)}$ interval. As one considers more perturbative contributions  both, the interval and $|M^2|$, decrease so it is plausible that at arbitrarily high orders the exact (and unique) critical value is the one  which satisfies  $g_c^{(w)} = g_c^{(s)} \equiv g_c$. 

Assuming that this is true one can use the low order results to estimate the critical coupling value at arbitrarily high orders  through a simple extrapolation (see right panel of Fig. \ref{Fig6}). By  interpolating over the four available (even) orders we  predict $g_c = 2.785$ as the extrapolated value. This simple exercise shows that although $M^2$ displays a much slower convergence than $M$ one can nevertheless use the information contained within the supercritical region to substantially improve the resummation of the physical mass squared.  Table \ref{tabela4} compares our best results with the most recent predictions for the critical coupling.

\begin{table}[ht!]
    \centering
    \begin{tabular}{ |c|c|c| }
        \hline
        Method & Year, Ref & $g_c$  \\
        \hline\hline
        Lattice Monte Carlo & 2009, \cite{lattice5} & $2.70^{+0.025}_{-0.013}$   \\ 
        Uniform matrix product states & 2013, \cite{lattice2} & 2.766(5)    \\
        Lattice Monte Carlo & 2015, \cite{lattice1} & 2.788(15)(8)  \\
        Resummed perturbation theory & 2015, \cite{rpt} & 2.75(1) \\
        \hline
        LO renormalized HT & 2015, \cite{hamilton1} & 2.97(14) \\
        raw HT & 2016, \cite{hamilton3} & 2.78(6) \\
        NLO-HT & 2017, \cite{hamilton7} & 2.76(3)\\
        \hline
        Borel resummation & 2018, \cite{serone1} & 2.807(34) \\
        \hline
        OPT optimal $M$ to ${\rm N}^7{\rm LO}$ & 2021, This work & 2.779(25)\\
        OPT optimal (extrapolated) $M^2$   & 2021, This work & 2.785\\
        \hline
    \end{tabular}
    \caption{Summary of recent predictions.}
    \label{tabela4}
    \end{table}

Before closing this section it is important to  emphasize  that one should not interpret our extrapolated value, $g_c =2.785$, as representing the exact one given that it has been obtained through a very simple numerical extrapolation procedure.  Rather, one should take this estimate as an indication that   the physical mass squared  
 can also be used to predict accurate critical coupling values provided that one uses the information stored within the supercritical region  to accelerate convergence.

\section{Conclusions}

We have applied  the OPT method to the bidimensional $\phi^4$ model in order to resum the ${\rm N}^8{\rm LO}$ perturbative series representing the mass gap within the $\ms$ renormalization scheme. Our main goal was to evaluate the numerical value of the critical coupling, $g_c$, at which a second order phase transition signals the  breaking of the $\mathbb{Z}_2$ symmetry. When the physical mass squared  is considered the OPT optimization criterion has real solutions only at even orders. In this case, the value $g_c = 1.511$ found at NNLO increases as higher orders are considered reaching the value $g_c = 2.373$ at ${\rm N}^8{\rm LO}$. Following Ref.  \cite {serone1} we have also resummed the series representing the mass, $M$, observing that  for this quantity the OPT optimization criterion has real solutions only at odd orders. Our investigation confirms that in this case  a better convergence is achieved: as higher orders are considered,  the value $g_c = 3.760$ found at NLO decreases to $g_c = 2.779$ at ${\rm N}^7{\rm LO}$. It turns out that the latter value is in excellent agreement with the state of the art ${\rm N}^8{\rm LO}$ result, $g_c = 2.807(34)$, obtained in Ref.  \cite {serone1} where $M$ was (Borel) resummed. 

Our prediction also agrees very well with results obtained with methods such as Hamiltonian truncation \cite{hamilton1, hamilton3, hamilton7}, lattice Monte Carlo \cite{lattice5, lattice1} and matrix product states \cite {lattice2} which predict $g_c$ to lie within the range $ 2.75 - 2.788$.  As   emphasized in the text, the logarithmic terms of the form $\ln (\mu^2/m^2)$ which are associated with tadpole contributions  must be included in the perturbative evaluation (prior to the OPT resummation) even when one later decides to set $\mu=m$. The reason is that within the OPT these terms become $\ln [\mu^2/(m^2+\eta^2)]$ so that even when $\mu=m$ the logarithms give important contributions as we have explicitly observed when resumming $M$ at  NLO. In this case, an order-$g$ logarithmic term was crucial to generate a highly non-perturbative variational mass allowing us to show that a decent prediction for $g_c$ can readily be obtained even when only a simple tadpole contribution is considered.  

Motivated by some recent observations \cite {serone4} we have investigated  the physical mass squared  behavior in the supercritical domain.  As a novelty, the present work  shows that this (often neglected)   region contains useful information which allows for high order extrapolations  to be performed.  Indeed, applying our extrapolation method to $M^2$ has substantially improved the  ${\rm N}^8{\rm LO}$ prediction generating values which are compatible with the OPT resummation of $M$, at ${\rm N}^7{\rm LO}$, as well as with other recent estimates. The present application demonstrates how a purely perturbative series can be efficiently resummed by the OPT procedure.

\acknowledgments
We thank the authors of Ref. \cite{serone1} for sending us some of their numerical data and for valuable comments. We are also grateful to  Jean-Lo\"{\i}c Kneur, Rudnei Ramos and Paul Romatschke   for related discussions. M.B.P. is  partially supported by Conselho Nacional de Desenvolvimento Cient\'{\i}fico e Tecnol\'{o}gico (CNPq-Brazil), Process No 303846/2017-8 and by Coordena\c c\~{a}o  de Aperfei\c coamento de Pessoal de  N\'{\i}vel Superior - (CAPES-Brazil)  -
Finance  Code  001. G.O.H. thanks Coordena\c c\~{a}o  de Aperfei\c coamento de Pessoal de  N\'{\i}vel Superior - (CAPES-Brazil) for a MSc scholarship.
This work has also been financed  in  part  by  Instituto  Nacional  de  Ci\^encia  e Tecnologia de F\'{\i}sica Nuclear e Aplica\c c\~{o}es  (INCT-FNA), Process No.  464898/2014-5.
 
\providecommand{\href}[2]{#2}\begingroup\raggedright\endgroup

\end{document}